\documentclass[prl, amsfonts, amssymb, amsmath, reprint, showkeys, twocolumn, twoside, nofootinbib, aps, superscriptaddress, eprint]{revtex4-2}
\usepackage{pgfplots,graphicx,mathrsfs,hyperref,cleveref,placeins,balance,braket,tikz,quantikz,makecell}

\pgfplotsset{compat=1.16}
\begin{document}

\title{Quantum State Preparation with Resolution Refinement}

\author{Scott Bogner}
\affiliation{Facility for Rare Isotope Beams and Department of Physics and Astronomy, Michigan State University, East Lansing, Michigan 48824, USA}

\author{Heiko Hergert}
\affiliation{Facility for Rare Isotope Beams and Department of Physics and Astronomy, Michigan State University, East Lansing, Michigan 48824, USA}

\author{Morten Hjorth-Jensen}
\affiliation{Department of Physics and Center for Computing in Science Education, University of Oslo, N-0316 Oslo, Norway}

\author{Ryan LaRose}
\affiliation{Department of Computational Mathematics, Science, and Engineering; Department of Electrical and Computer Engineering; Center for Quantum Computing, Science, and Engineering; Department of Physics and Astronomy, Michigan State University, East Lansing, MI 48824, USA}

\author{Dean Lee}
\affiliation{Facility for Rare Isotope Beams and Department of Physics and Astronomy, Michigan State University, East Lansing, Michigan 48824, USA}

\author{Matthew Patkowski}
\affiliation{Department of Physics, University of Colorado Boulder, Boulder CO 80309, USA}

\begin{abstract} 
We introduce a method called resolution refinement that allows one to bootstrap eigenstate preparation on a quantum computer.  We first prepare an eigenstate of a low-resolution Hamiltonian using any method of choice.  The eigenstate is then lifted to higher resolution and adiabatically evolved to produce the corresponding eigenstate of a higher-fidelity Hamiltonian.  We give examples of resolution refinement applied to both single-particle basis states as well as a spatial lattice grid.  For basis refinement, we compute few-body ground states of the Busch model for interacting particles in a harmonic trap in one dimension.  For lattice refinement, we compute Hartree-Fock nuclear states for a central Woods-Saxon potential in three dimensions, and we compute bound states and continuum states in a multi-species Hubbard model of fermions in one dimension.  In all cases, the method is efficient and requires an adiabatic evolution time that scales with the inverse of the energy gap times the square root of the system size.  We show that this very favorable scaling arises from the fact that resolution refinement does not make large changes to the structure or energies of the low-energy eigenstates.
\end{abstract}

\maketitle

A central challenge for studying quantum many-body systems on a quantum computer is the efficient preparation of complex and correlated ground states.  This is a task for which many prominent algorithms face difficult scaling limits as the system size increases. Variational methods such as variational quantum eigensolvers \cite{peruzzo2014variational} can face the challenge of exponentially vanishing gradients for large systems \cite{mcclean2018barren}.  The runtime for adiabatic state preparation increases exponentially with system size if the initial and final Hamiltonians have significantly different ground state wavefunctions, due to hindered quantum tunneling through energy barriers \cite{albash2018adiabatic}. Filtering methods such as quantum phase estimation \cite{kitaev1995quantum,abrams1999quantum} and the rodeo algorithm \cite{Choi:2020pdg} are probabilistic in nature. Their success probabilities are proportional to the initial state overlap with the eigenstate of interest, and these decrease exponentially with system size.

Despite these challenges, many quantum algorithms are indeed efficient and accurate when applied to systems restricted to a small model space, such as a limited number of single-particle orbitals or a coarse lattice grid. It would therefore be useful to have some means of bootstrapping these low-resolution eigenstates to produce the corresponding eigenstates for a larger model space.  In this paper, we present a general approach called resolution refinement that performs this bootstrapping.  The first step is to prepare an eigenstate of a low-resolution Hamiltonian using any method of choice.  This is made possible because the model space is small. The eigenstate is then lifted to higher resolution and adiabatically evolved to produce the corresponding eigenstate of a higher-fidelity Hamiltonian.  We show that many-body calculations using products of single-particle orbital states can be bootstrapped using basis refinement.  We also show that the resolution refinement can be applied to many-body lattice Hamiltonian simulations using lattice refinement.  

Effective field theory and its implementation using renormalization group methods have been central to recent progress in nuclear theory using classical computation \cite{Bogner:1999my,Bogner:2006pc,Tsukiyama:2010rj,Hergert:2012nb,Tsunoda:2013bla,Hagen:2015yea,Ekstrom:2015rta,Elhatisari:2016owd,Stroberg:2016ung,Yao:2019rck,Lee:2020esp}.  Resolution refinement can be viewed as a physics-inspired quantum computing algorithm based on principles of low-energy effective field theory and renormalization group flow.  It acts like an inverse renormalization group transformation from low resolution to high resolution.  

As a first example, we consider the Busch model \cite{busch1998two} in one spatial dimension for $K$ distinguishable fermions with mass $m$ in a harmonic oscillator trap with angular frequency $\omega$ and same-strength delta function interactions between all pairs.  We use harmonic oscillator units where length is in units of $\sqrt{\hbar/m\omega}$ and energy is in units of $\hbar \omega$.  In the formalism of second quantization with creation and annihilation operators $a^\dagger_{x,k}$ and $a_{x,k}$, the Hamiltonian $H$ has the form,
\begin{equation}
   \tfrac{-1}{2}\sum_k\int dx \,a^\dagger_{x,k}\nabla^2 a_{x,k} + \tfrac{1}{2}\int dx \left[x^2\rho_{x}+g:\rho^2_{x}:\right],
\end{equation}
where $\rho_{x}$ is the local total density $\sum_k a^\dagger_{x,k}a_{x,k}$.  The $::$ symbol denotes normal ordering, where annihilation operators are on the right and creation operators on the left. 
We use single-particle orbitals corresponding to harmonic oscillator eigenstates with quantum numbers $n=0,\cdots,n_{\rm max}-1$, where $n_{\rm max}$ is the basis truncation parameter.  On a quantum computer, we assign one qubit to each of the $Kn_{\rm max}$ modes.

We consider basis refinement calculations performed at low resolution and $n_{\rm max}=2$ and $n_{\rm max}=10$.  We let $N$ be the total number of fermions.  For $K=N=2$, the ground state energy is known analytically \cite{busch1998two}, and the value is $1.3067$ for coupling $g=1$.  At low resolution with $n_{\rm max}=2$, the ground state energy is $1.3782$.  At high resolution with $n_{\rm max}=10$, the energy is $1.3264$.  Let $H_{\rm low}$ be the low-resolution Hamiltonian for the $n_{\rm max}=2$ truncation.  We define a prolongation operator $P$ that lifts every many-body basis state in the low-resolution space to the same many-body basis state in the $n_{\rm max}=10$ high-resolution space.  The restriction operator $P^\dagger$ performs the projection from high-resolution space to the low-resolution space.  Since the kernel of $P^\dagger$ is large, we use an energy shift $\mu$ large enough to ensure the low-energy states of interest have energies below zero, well separated from the null space. This process can be implemented using first quantization with matrix quantum mechanics or second quantization with creation and annihilation operators. 

The resolution refinement process consists of performing adiabatic evolution with time duration $T$ from the shifted and prolonged low-resolution Hamiltonian $P(H_{\rm low}-\mu)P^\dagger$ to the shifted high-resolution Hamiltonian $H_{\rm high}-\mu$.  We shift all the energies back by an amount $\mu$ afterwards.  For all of the adiabatic evolution calculations in this work we use the interpolation functions $\cos^2 [\theta(t)]$ and $\sin^2 [\theta(t)]$ for the coefficients of $P(H_{\rm low}-\mu)P^\dagger$ and $H_{\rm high}-\mu$, respectively, with $\theta(t)=(\pi t)/(2T)$ and $t$ ranging from $0$ to $T$.  This process is the same for basis refinement and for lattice refinement.  

While resolution refinement can be applied to any low-energy eigenstate, we focus on ground state calculations in the present work.  We can quantify the fidelity of resolution refinement by computing the overlap probability $\lvert\braket{\Psi_0|\Phi(T)}\rvert^2$, where $\ket{\Phi(T)}$ is the final wavefunction after adiabatic evolution and $\ket{\Psi_0}$ is the ground state of $H_{\rm high}$.  

\begin{figure}
    \centering
    \includegraphics[width=0.9\linewidth]{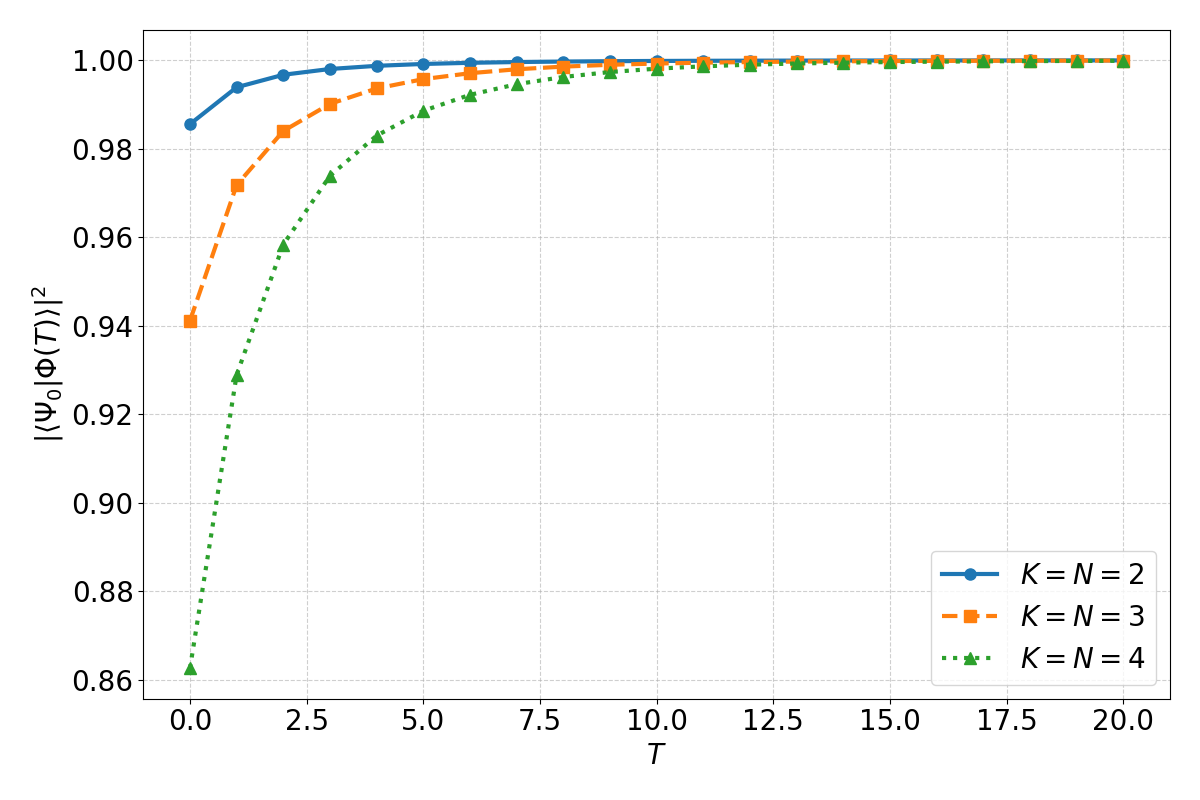}
    \caption{Plot of the ground state overlap probabilities versus total adiabatic evolution time for $K=N=2$, $K=N=3$, and $K=N=4$ for the Busch model in one dimension.}
    \label{fig:overlap_Busch}
\end{figure}

In Fig.~\ref{fig:overlap_Busch}, we plot the ground state overlap probability versus total adiabatic evolution time $T$ for $K=N=2$, $K=N=3$, and $K=N=4$.  For each case, the low-resolution calculation uses $n_{\rm max}=2$ and the high-resolution calculation uses $n_{\rm max}=10$.  The energy gap, $\Delta E$, for $K=N=2$ is about $0.7$.  The energy gap is about $0.6$ for $K=N=3$ and about $0.5$ for $K=N=4$. We see that the overlap probabilities approach $1$ with timescale $T \sim (\Delta E)^{-1}$ that grows slowly with $N$.  The convergence of adiabatic evolution is typically exponential at early times, but switches over to power law convergence at late times.  As discussed in Ref.~\cite{Patkowski:2025leg}, it is therefore useful to combine adiabatic evolution with filtering algorithms such as the rodeo algorithm in order to have exponential convergence for late times also.

\begin{figure}
    \centering
    \includegraphics[width=0.6\linewidth]{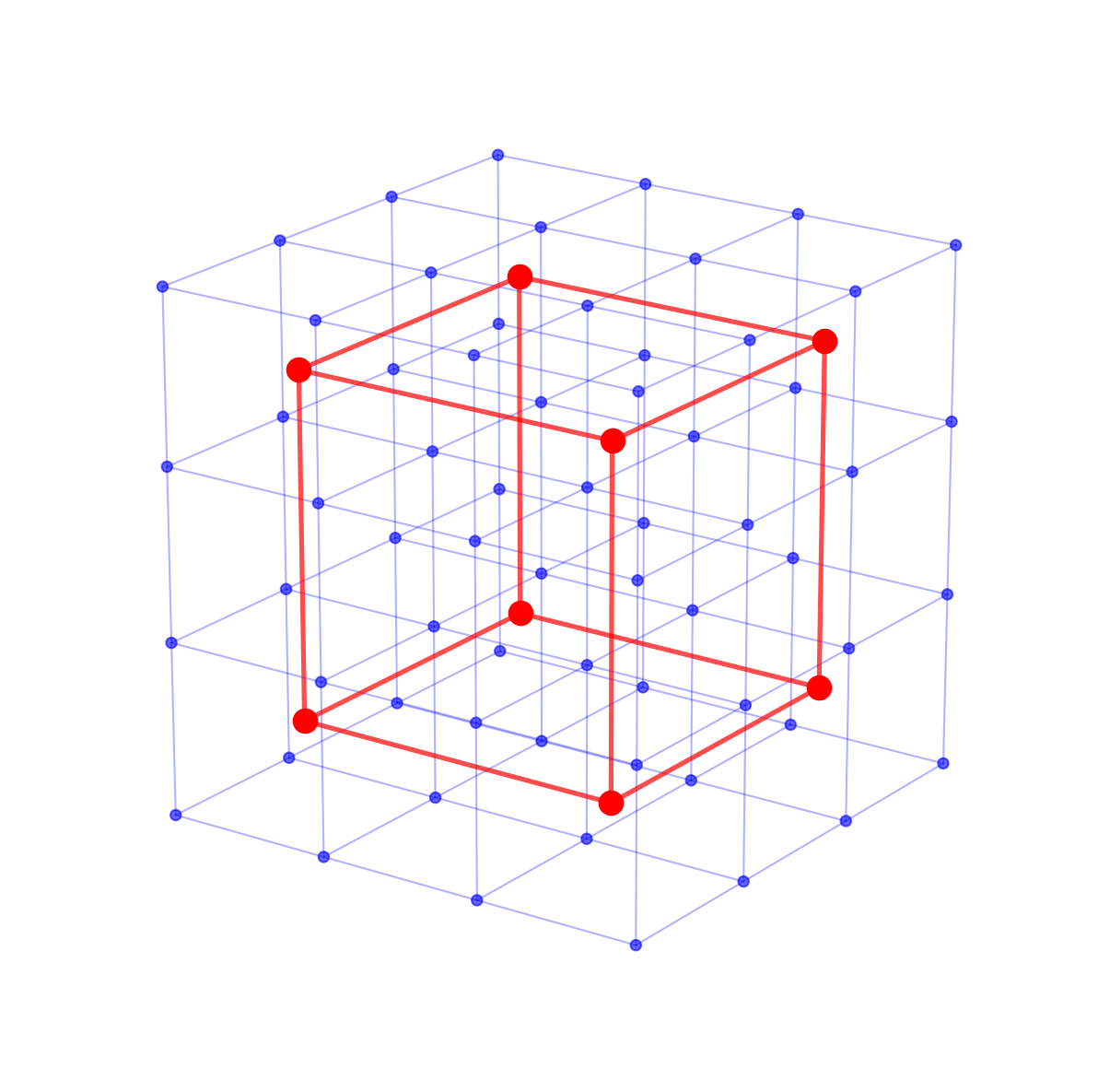}
    \caption{Sketch of the fine lattice grid with spacing $a$ and the coarse lattice grid with spacing $2a$.}
    \label{fig:lattice}
\end{figure}

We now turn our attention to lattice refinement for lattice Hamiltonians.  For lattice calculations on a quantum computer, we use the second quantization formalism with the occupation basis for each particle species, where $\ket{0}$ or $\ket{1}$ represents a particle of that species occupying a given lattice site or not.  For lattice refinement we need a prolongation operator $P$ that lifts many-body states on a coarse lattice with spacing $2a$ to the corresponding many-body states on a fine lattice grid with spacing $a$.  The two lattice grids are illustrated in Fig.~\ref{fig:lattice} for a three-dimensional lattice.  Any operator at coarse lattice site $2\boldsymbol{r}$ becomes a normalized sum of operators at surrounding fine lattice sites $2\boldsymbol{r}\pm\boldsymbol{\hat{n}}/2$, where $\boldsymbol{\hat{n}}$ are lattice unit vectors along the $d$ spatial directions.

\begin{figure}[h!]
    \centering 
    \begin{quantikz}
        \lstick{$\ket{q_1}$} & \qw      & \ctrl{1} & \qw      & \targ{}   & \qw \\
        \lstick{$\ket{q_2}$} & \gate{R_y(\frac{\pi}{4})} & \targ{}  & \gate{R_y(-\frac{\pi}{4})} & \ctrl{-1} & \qw
    \end{quantikz}
    \caption{The circuit decomposition for the unitary $U$ operator built from single-qubit rotations and two CNOT gates.}
    \label{fig:circuit}
\end{figure}
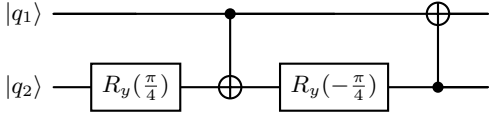

To implement the prolongation operator on a quantum computer, we distribute any $\ket{1}$ state on a coarse lattice site to an equal superposition of $2^d$ states with exactly one $\ket{1}$ state on the surrounding fine lattice sites.  For any $\ket{0}$ state on a coarse lattice site, we set all the surrounding fine lattice sites to $\ket{0}$.  We first illustrate the process for one spatial dimension.  For each qubit corresponding to a coarse lattice site, we introduce a second qubit initialized in the $\ket{0}$ state.  We then apply a unitary operation $U$ to these two qubits where $U\ket{00} = \ket{00}$ and $U\ket{10} = \frac{1}{\sqrt{2}}(\ket{10} + \ket{01})$. 
Fig.~\ref{fig:circuit} shows a simple circuit that performs the required operations using two single-qubit rotations and two CNOT gates.
This circuit provides the required prolongation along the $x$ direction.  For more than one spatial dimension, we simply apply the same process recursively for each spatial direction to the qubits obtained in the previous prolongation step.  

This approach is suitable for lattice simulations of many-body fermions and hard core bosons.  For fermionic particles, it is convenient to choose a canonical ordering of modes such that each of the $2^d$ qubits are contiguous in the ordering.  This convenient choice eliminates the need for keeping track of minus signs during the prolongation step.  When performing adiabatic evolution, we break up the Trotter product with exponentials of terms for $P(H_{\rm low}-\mu)P^\dagger$ and exponentials of terms for $H_{\rm high}-\mu$.  The exponentials of terms for $P(H_{\rm low}-\mu)P^\dagger$ can be performed efficiently using fermionic fast Fourier transforms for each $2^d$ cube for fermions \cite{Verstraete:2008qpa,Babbush:2018zma} and quantum fast Fourier transforms for hard core bosons \cite{Somma:2001kjh}.  For lattice gauge theories, there are additional complications due to gauge fields residing on the lattice links as well as the Gauss' law constraints.  The successful implementation of lattice refinement for gauge theories is beyond the scope of the present work.

For our benchmark lattice calculations, we consider a Hubbard model with $K$ fermion species with mass $m$ on a periodic cubic lattice in $d$ spatial dimensions.  We use natural units where $\hbar$ and the speed of light, $c$, are set to $1$.  The Hamiltonian $H$ has the form
\begin{equation}
    \tfrac{-1}{2mb^2}\!\!\sum_{\langle \boldsymbol{r}, \boldsymbol{r}' \rangle,k}
    \! a_{\boldsymbol{r},k}^\dagger a_{\boldsymbol{r}'\!,k} 
    + \! \sum_{\boldsymbol{r},j<k}  \!C_{j,k}\rho_{\boldsymbol{r},j}\rho_{\boldsymbol{r},k}
    + \sum_{\boldsymbol{r},k} V_{\boldsymbol{r}} \rho_{\boldsymbol{r},k},
    \label{eq:Hfine}
\end{equation}
where $b$ is the lattice spacing, $\langle \boldsymbol{r}, \boldsymbol{r}'\rangle$ denotes nearest neighbors, and 
$\rho_{\boldsymbol{r},k}$ is the local density operator 
$a_{\boldsymbol{r},k}^\dagger a_{\boldsymbol{r},k}$ for species $k$. In the single-particle potential $V_{\boldsymbol{r}}$, we have included the term $d/(mb^2)$ needed to complete the finite-difference Laplacian in $d$ dimensions.  The parameters $C_{j,k}$ control the two-body interactions. 

For the first lattice benchmark, we set $C_{j,k}=0$ and calculate Hartree-Fock lattice wavefunctions for a central Woods-Saxon potential of the form \begin{equation}
    V_{\boldsymbol{r}}=-\frac{V_0}{1+e^{(|\boldsymbol{r}|-R)/\alpha}}, \; R = R_0A^{1/3},
\end{equation}
where $A$ is the number of nucleons.  The number of nucleon species is $K=4$. For the fine lattice grid, we take $a=(150~{\rm MeV})^{-1}=1.32~{\rm fm}$, matching recent nuclear lattice effective field theory calculations \cite{Lu:2021tab,Shen:2022bak,Elhatisari:2022zrb,Ma:2023ahg,Meissner:2023cvo,Giacalone:2024luz,Giacalone:2024ixe,Konig:2023rwe}.  The coarse lattice grid then has lattice spacing $2a=(75~{\rm MeV})^{-1}=2.63~{\rm fm}$.  The nucleon mass is $m = 938.92~{\rm MeV}$, and the diffuseness parameter is taken to be $\alpha=0.5~{\rm fm}$.  

For the depth $V_0$ and radius parameter $R_0$, we adjust the parameters for the coarse and fine lattice Hamiltonians so that they produce similar energies and radii.  We take $R_0=1.5~{\rm fm}$ and $V_0=50~{\rm MeV}$ for the fine lattice grid, and we take $R_0=1.8~{\rm fm}$ and $V_0=40~{\rm MeV}$ for the coarse lattice grid.  For the lattice calculations, we need slightly larger values for $R_0$ in order to produce nuclear radii and binding energies similar to those obtained in continuous space.

\begin{figure}
    \centering
    \includegraphics[width=0.9\linewidth]{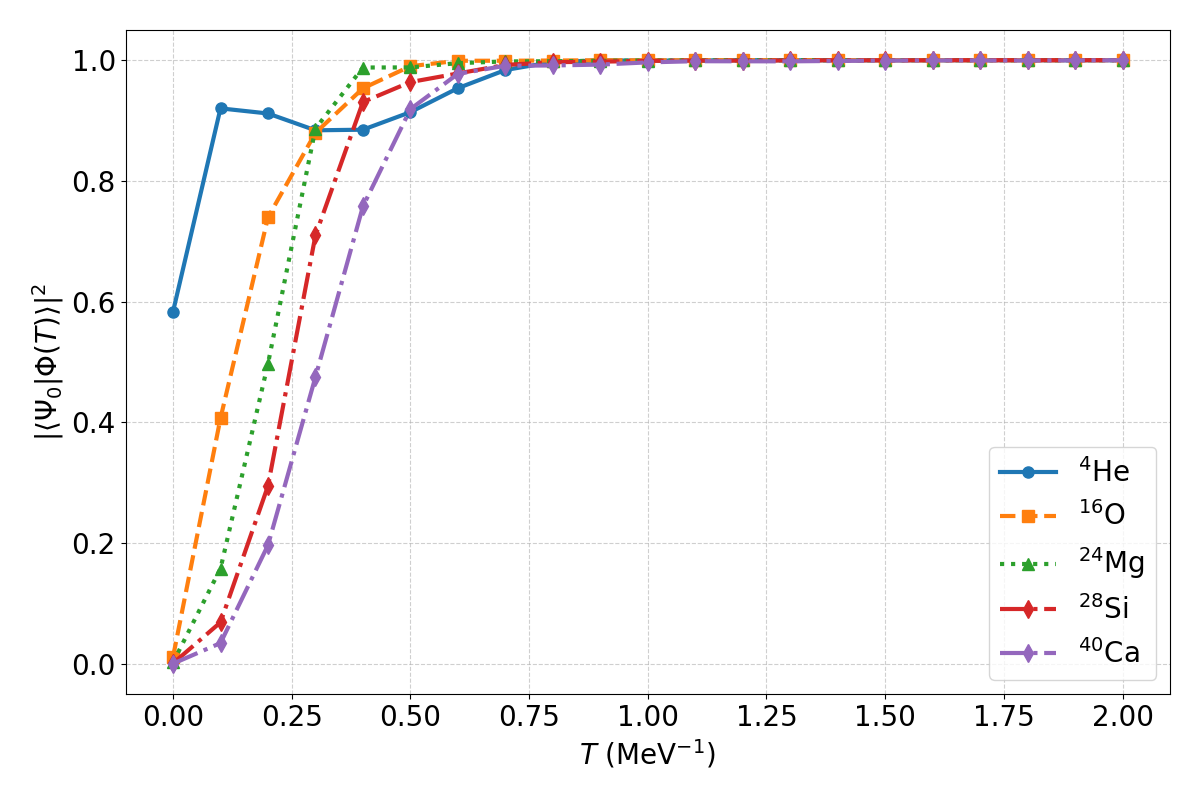}
    \caption{Plot of the overlap probabilities versus total adiabatic evolution time for $^4$He, $^{16}$O, $^{24}$Mg, $^{28}$Si and $^{40}$Ca.}
    \label{fig:overlap_WS}
\end{figure}

\begin{figure}
    \centering
    \includegraphics[width=0.9\linewidth]{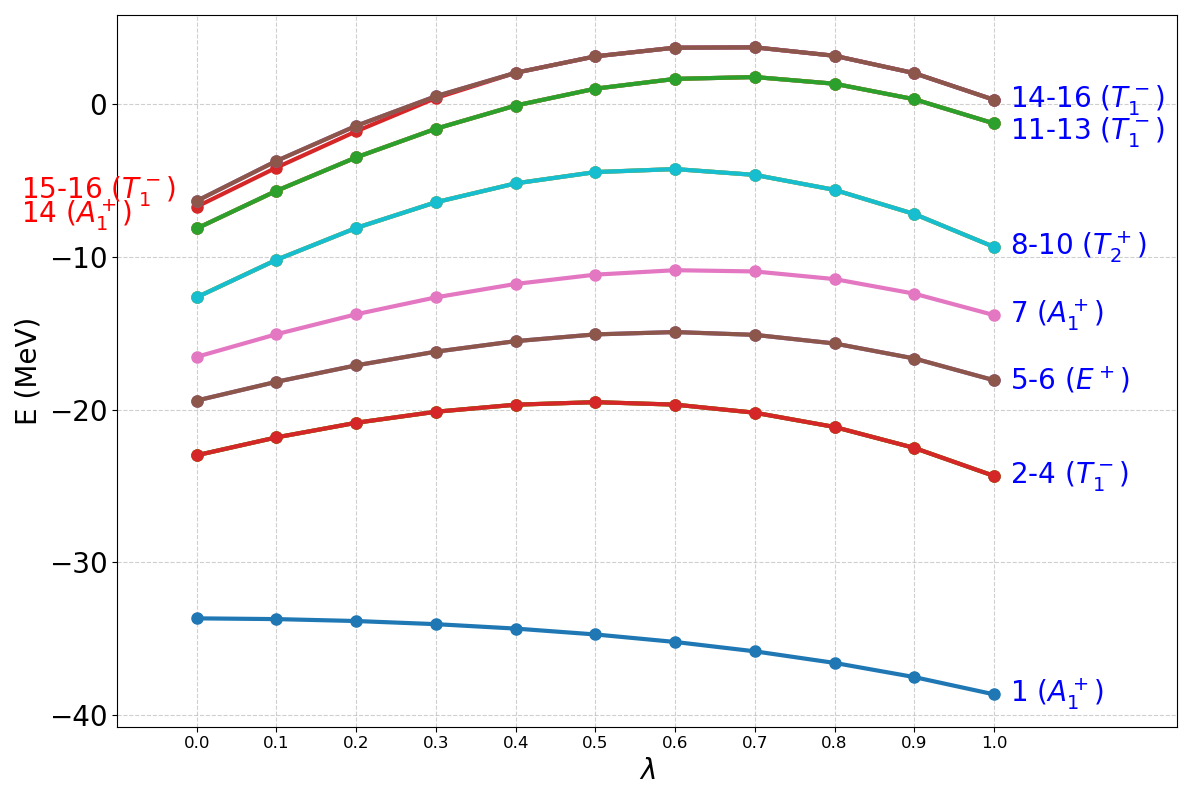}
    \caption{Single-particle energy levels for one nucleon species for the $^{40}$Ca calculation as a function of interpolation parameter $\lambda$.  We have labeled the corresponding irreducible cubic representations.}
    \label{fig:energies_A40}
\end{figure}

We use lattice refinement to compute the ground state wavefunctions of $^4$He, $^{16}$O, $^{24}$Mg, $^{28}$Si, and $^{40}$Ca.  For each of these calculations, the energy gap $\Delta E$ is between $10$ and $15$ MeV.  In Fig.~\ref{fig:overlap_WS} we show the overlap probability $\lvert\braket{\Psi_0|\Phi(T)}\rvert^2$ versus total evolution time $T$.  We see that the overlap probability approaches $1$ with a timescale that scales as $(\Delta E)^{-1}$ and grows slowly with the number of particles.  For $^{40}$Ca, the overlap probability starts at $8\times 10^{-6}$ and rises by five orders of magnitude.  For $^4$He there are some oscillations as often happens in adiabatic evolution.  This performance can be improved by using adiabatic evolution for $T \approx 0.1~{\rm MeV}^{-1}$ and then switching to a filtering method such as the rodeo algorithm to get exponential convergence of the overlap probability at later times.
The fact that the lattice refinement is not producing any difficulties can be seen by plotting the energy levels for $H_\lambda$, which linearly interpolates between $P(H_{\rm low}-\mu)P^\dagger+\mu$ at $\lambda=0$ and $H_{\rm high}$ at $\lambda=1$.  In Fig.~\ref{fig:energies_A40}, we plot the single-particle energy levels for $H_\lambda$ for one nucleon species in $^{40}$Ca.  We have labeled the corresponding irreducible cubic representations.  We see that the curves are very smooth, making the adiabatic evolution efficient.

For our last benchmark calculation, we compute the five-body bound states and continuum states for $K=5$ distinguishable fermions in the one-dimensional Hubbard model.  We again take the fermion masses to be $m = 938.92~{\rm MeV}$, and we take $V_{\boldsymbol{r}}$ to contain the Laplacian term only, $1/(mb^2)$.  In Table~\ref{tab:coefficient}, we show the two-body interaction coefficients used to produce cluster ground states with $1+1+1+1+1$, $2+1+1+1$, $2+2+1$, $3+1+1$, $3+2$, $4+1$, and $5$ particles.

\begin{table}[htbp]
        \caption{Two-body interaction coefficients producing the various cluster ground states.}    
        \label{tab:coefficient}
    \begin{tabular}{|c|c|c|c|}
        \hline
        state & $C_{j,k}=-20~{\rm MeV}$ & $C_{j,k}=10~{\rm MeV}$ & $C_{j,k}=20~{\rm MeV}$\\
        \hline
        \makecell{1+1+1 \\+1+1} & & \makecell{(1,5), (2,3),\\(3,5), (4,5)} & 
        \makecell{(1,2), (2,4), \\(3,4)}\\
        \hline      
        \makecell{2+1+1 \\+1} & (1,2) & \makecell{(1,5), (2,3),\\(3,5), (4,5)} & 
        (2,4), (3,4)\\
        \hline          
        2+2+1 & (1,2), (3,4) & \makecell{(1,5), (2,3),\\(3,5), (4,5)} & 
        (2,4)\\  
        \hline          
        3+1+1 & (1,2), (2,3) & (1,5) & (2,4)\\        
        \hline              
        3+2 & \makecell{(1,2), (2,3), \\(4,5)} &  & (2,4)\\
        \hline
        4+1 & \makecell{(1,2), (2,3), \\(2,4)} & & (4,5) \\
        \hline   
        5 & \makecell{(1,2), (2,3), \\(2,4), (4,5)} & & \\
        \hline
    \end{tabular}
\end{table}

\begin{figure}
    \centering
    \includegraphics[width=0.9\linewidth]{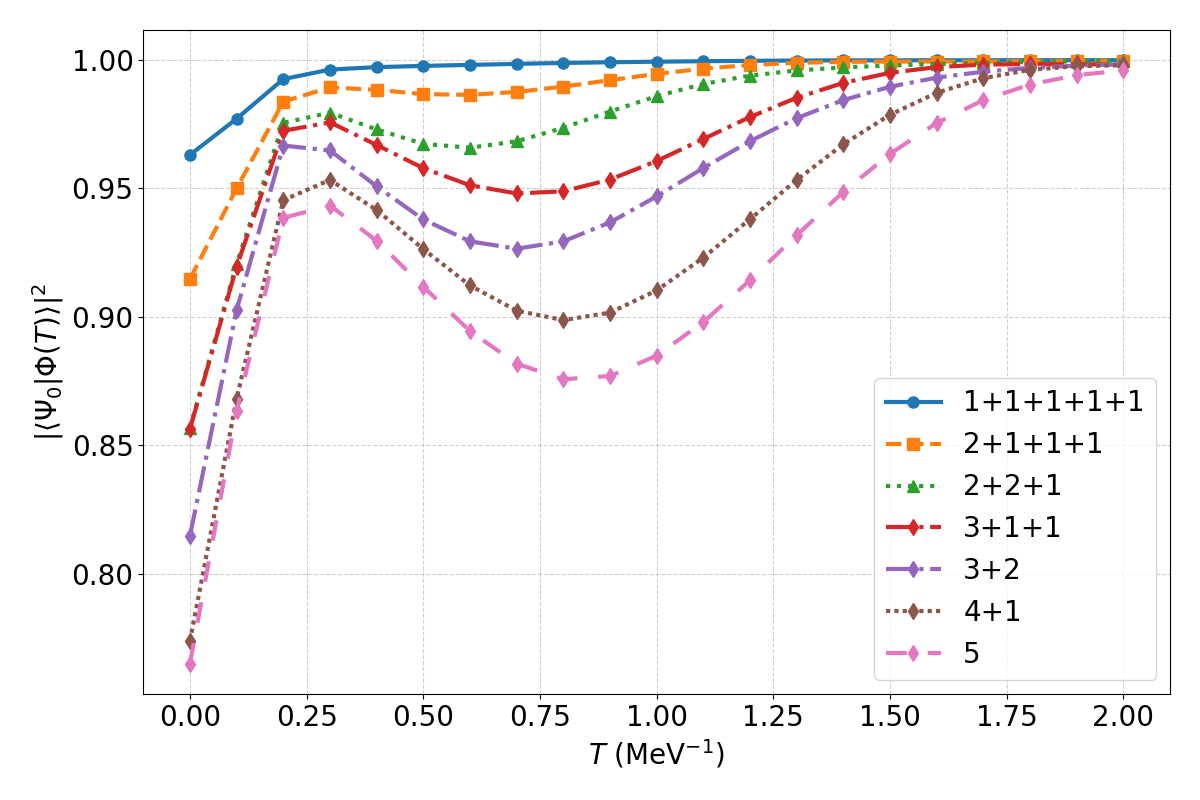}
    \caption{Ground state overlap probability for cluster states composed of $1+1+1+1+1$, $2+1+1+1$, $2+2+1$, $3+1+1$, $3+2$, $4+1$, and $5$ particles for various one-dimensional Hubbard model Hamiltonians. We plot $|\braket{\Psi_0|\Phi(T)}|^2$ versus total adiabatic evolution time $T$.}
    \label{fig:overlap_cluster}
\end{figure}

 In Fig.~\ref{fig:overlap_cluster} we show the overlap probability $\lvert\braket{\Psi_0|\Phi(T)}\rvert^2$ versus total evolution time $T$ for the various one-dimensional Hubbard model Hamiltonians corresponding to the interaction coefficients in Table~\ref{tab:coefficient}.  If we do not count excited states that are translational modes of the ground states, the energy gap $\Delta E$ for each Hamiltonian considered is about $4$ or $5$ MeV.  As found before, the overlap probability approaches $1$ with a timescale proportional to $(\Delta E)^{-1}$ and growing slowly with $N$.  There are again oscillations at later times, but as before we can limit adiabatic evolution to $T \approx 0.25~{\rm MeV}^{-1}$ and switch to a filtering method to complete the state preparation.

 In this work, we have presented a technique called resolution refinement that allows one to bootstrap eigenstate preparation.  We start with an eigenstate of a low-resolution Hamiltonian and lift the eigenstate to a high-resolution space.  Adiabatic evolution is then used to prepare the corresponding eigenstate of a high-resolution Hamiltonian.  Filtering methods can be used to augment the performance of adiabatic evolution.  We have presented benchmark examples for basis refinement as well as lattice refinement.  In all cases, the adiabatic evolution time required is comparable to the inverse of the energy gap in the physical spectrum.  This scaling of the adiabatic time is much smaller than the $O(1/(\Delta E_{\rm min})^2)$ scaling for most applications of adiabatic evolution, where $\Delta E_{\rm min}$ is the minimum energy gap along the adiabatic path.  
 
 The large reduction in computational effort arises from the fact that resolution refinement does not make large changes to the structure or energies of the low-energy eigenstates.  This means that $\Delta E_{\rm min}$ remains comparable to the energy gap $\Delta E$ in the physical spectrum.  Furthermore, the loss of fidelity during adiabatic evolution at large times is directly connected to the squared norm of the first-order correction to the wavefunction when treating the difference between final and initial Hamiltonians as a perturbation.  If the squared norm of this first-order correction does not become large, then the required adiabatic evolution time scales with the inverse of the energy gap. In the Supplemental Material, we show that the adiabatic time $T$ scales as $O\left(\frac{\sqrt{N}}{\sqrt{\epsilon}\Delta E_{\rm min}}\right)$ when using linear interpolation for the adiabatic path, where $N$ is the number of particles and the target fidelity is $1-\epsilon$.
 
 Resolution refinement is a general purpose and practical approach based on the principles of effective field theory that can be used to extend the reach of currently used many-body state preparation methods. While the method may fail for systems that are extremely large or manifest correlations that are not properly captured by the low-resolution Hamiltonian, it appears broadly useful for a wide range of quantum many-body problems. 

{\it We acknowledge financial support from the U.S. Department of Energy through grants DE-SC0023658, DE-SC0013365, DE-SC0024586, DE-SC0023175, DE-SC0026198, DE-SC0023516, and the U.S. National Science Foundation through grants PHY-2310620 and PHY-2310020.  This research used resources of the Oak Ridge Leadership Computing Facility, which is a DOE Office of Science User Facility supported under contract DE-AC05-00OR22725.}
\bibliographystyle{apsrev4-1} 
\bibliography{References}


\clearpage

\setcounter{page}{1}
\renewcommand{\thepage}{S\arabic{page}}
\setcounter{equation}{0}
\renewcommand{\theequation}{S\arabic{equation}}
\setcounter{figure}{0}
\renewcommand{\thefigure}{S\arabic{figure}}
\setcounter{table}{0}
\renewcommand{\thetable}{S\arabic{table}}

\onecolumngrid
\begin{center}
  \textbf{\large Supplemental Material}
\end{center}

\section{Derivation of Adiabatic Evolution Time Scaling}

When using adiabatic evolution to interpolate between $P(H_{\rm low}-\mu)P^\dagger$ and $H_{\rm high}-\mu$, we have observed empirically that the adiabatic evolution time $T$ needed scales inversely with the minimum energy gap $\Delta E_{\rm min}$ and grows slowly with system size. In this section, we provide a detailed derivation of this observed scaling for the case of linear interpolation for the adiabatic path.  The discussion focuses on adiabatic preparation of the ground state.  However, the same derivation applies to any non-degenerate eigenstate for any chosen set of quantum  numbers.  For the case of excited states, the energy gap refers to the gap in the spectrum above and below the chosen energy eigenstate.

\subsection{Statement of Result}

\textbf{Proposition.} \textit{Let $H(s)$ for $s \in [0,1]$ be the adiabatic path interpolating between the low-resolution Hamiltonian $H(0)=P(H_{\mathrm{low}}-\mu)P^\dagger$ and the high-resolution Hamiltonian $H(1)=H_{\mathrm{high}}-\mu$. In order to simplify the analysis, we consider linear interpolation where $H(s) = (1-s) H(0) + s H(1)$.  Let $|\Psi_0(s)\rangle$ be the instantaneous ground state, and let $|\Psi_n(s)\rangle$ for $n \ne 0$ label the instantaneous excited states with the same quantum numbers as the ground state.  The corresponding instantaneous energies are $E_0(s)$ and $E_n(s)$.  We assume a non-vanishing spectral gap $\Delta_{n0}(s) \equiv E_n(s) - E_0(s) > 0$ for all $n \ne 0$. Treating $V = H(1) - H(0)$ as a perturbation, we define $\mathcal{N}_{\mathrm{pert}}(s)$ as the squared norm of the first-order wavefunction correction,}
\begin{equation}
    \mathcal{N}_{\mathrm{pert}}(s) \equiv \sum_{n \neq 0} \frac{|\langle \Psi_n(s) | V | \Psi_0(s) \rangle|^2}{(\Delta_{n0}(s))^2}.
\end{equation}
\textit{We are using the intermediate normalization convention where all perturbative corrections are orthogonal to the unperturbed wavefunction. If $\mathcal{N}_{\mathrm{pert}}(s)$ is bounded above by a constant $B$,}
\begin{equation}
   B \equiv \max_{s \in [0,1]} \mathcal{N}_{\mathrm{pert}}(s) ,
\end{equation}
\textit{and $\Delta_{n0}(s)$ is bounded below by $\Delta E_{\rm min}$,}
\begin{equation}
  \Delta E_{\rm min} \equiv \min_{s \in [0,1]}  \Delta_{n0}(s),
\end{equation}
\textit{then the fidelity loss $P_{\mathrm{loss}}$ is bounded above by}
\begin{equation}
    P_{\mathrm{loss}} \le \frac{4 B}{T^2 (\Delta E_{\rm min})^2} + \mathcal{O}(T^{-3}),
\end{equation}
\textit{and the adiabatic evolution time $T$ required to ensure a final ground state fidelity $\mathcal{F}(T) \ge 1-\epsilon$ scales asymptotically as}
\begin{equation}
    T \sim \mathcal{O}\left( \frac{\sqrt{B}}{\sqrt{\epsilon} \Delta E_{\rm min}} \right).
\end{equation}

\subsection{Proof}

\subsubsection{Definitions and Dynamics}
We define the fidelity of the state preparation as the overlap probability between the time-evolved state $|\Phi(T)\rangle$ and the target ground state $|\Psi_0(1)\rangle$,
\begin{equation}
    \mathcal{F}(T) = |\langle \Psi_0(1) | \Phi(T) \rangle|^2.
\end{equation}
The loss of fidelity, or the probability of finding the system in any excited state, is $P_{\mathrm{loss}} = 1 - \mathcal{F}(T)$.

The time-evolved state $|\Phi(s)\rangle$ satisfies the Schrödinger equation $i T^{-1} \partial_s |\Phi(s)\rangle = H(s) |\Phi(s)\rangle$. We expand this state in the instantaneous eigenbasis $\{|\Psi_n(s)\rangle\}$ with the dynamical phase factored out \cite{Teufel:2003},
\begin{equation}
    |\Phi(s)\rangle = \sum_n c_n(s) \exp\left( -i T \int_0^s E_n(\tau) d\tau \right) |\Psi_n(s)\rangle.
\end{equation}
We assume a parallel-transported basis where the geometrical phases have been absorbed into the definitions of the eigenstates, such that $\langle \Psi_n(s) | \partial_s \Psi_n(s) \rangle = 0$. We assume the system is initialized in the ground state of the low-resolution Hamiltonian, and so the initial conditions are $c_0(0) = 1$ and $c_{n \neq 0}(0) = 0$.

\subsubsection{Transition Amplitudes}
Substituting the basis expansion into the Schrödinger equation yields a system of coupled differential equations for the coefficients $c_n(s)$. We first differentiate the ansatz for $|\Phi(s)\rangle$ with respect to the scaled time parameter $s$
\begin{equation}
    \partial_s |\Phi(s)\rangle = \sum_k e^{-iT \int_0^s E_k(\tau) d\tau} \left[ (\partial_s c_k) - i T E_k c_k + c_k \partial_s \right] |\Psi_k(s)\rangle.
\end{equation}
Multiplying by $i T^{-1}$ and substituting into the Schrödinger equation $i T^{-1} \partial_s |\Phi(s)\rangle = H(s)|\Phi(s)\rangle$ leads to the cancellation of the energy terms proportional to $E_k$. This simplification results in the equation
\begin{equation}
    \sum_k e^{-iT \int_0^s E_k(\tau) d\tau} \left[ (\partial_s c_k) |\Psi_k(s)\rangle + c_k |\partial_s \Psi_k(s)\rangle \right] = 0.
\end{equation}
We project this equation onto the instantaneous excited state $\langle \Psi_n(s)|$. Utilizing the orthonormality condition $\langle \Psi_n(s) | \Psi_k(s) \rangle = \delta_{nk}$ allows us to isolate the derivative of the coefficient $c_n$,
\begin{equation}
    \partial_s c_n(s) = - \sum_k c_k(s) \langle \Psi_n(s) | \partial_s \Psi_k(s) \rangle e^{i T \int_0^s (E_n(\tau) - E_k(\tau)) d\tau}.
\end{equation}
The sum over $k$ includes contributions from the ground state and all excited states. We assume the basis is parallel-transported such that $\langle \Psi_n(s) | \partial_s \Psi_n(s) \rangle = 0$, which eliminates the diagonal $k=n$ term. In the adiabatic limit for large $T$, the system remains primarily in the ground state, implying $c_0(s) \approx 1$ and $c_{k \neq 0}(s) \ll 1$. Neglecting transitions between excited states, which are suppressed by higher powers of $1/T$, the sum is dominated by the $k=0$ term~\cite{albash2018adiabatic},
\begin{equation}
    \partial_s c_n(s) = - \langle \Psi_n(s) |  \partial_s \Psi_0(s) \rangle \exp\left( i T \int_0^s \Delta_{n0}(\tau) d\tau \right) + \mathcal{O}(T^{-1}).
    \label{eq:deriv_c}
\end{equation}
To determine the geometric coupling matrix element $\langle \Psi_n | \partial_s \Psi_0 \rangle$, we differentiate the instantaneous eigenvalue equation $H(s)|\Psi_0(s)\rangle = E_0(s)|\Psi_0(s)\rangle$ with respect to $s$. Projecting the result onto the excited state $\langle \Psi_n(s)|$ and using the Hermiticity of $H(s)$ yields the relation
\begin{equation}
    \langle \Psi_n(s) | (\partial_s H(s)) | \Psi_0(s) \rangle + E_n(s) \langle \Psi_n(s) | \partial_s \Psi_0(s) \rangle = E_0(s) \langle \Psi_n(s) | \partial_s \Psi_0(s) \rangle.
\end{equation}
Solving for the overlap term and identifying $\partial_s H(s)$ with $V$, assuming a linear interpolation path, we obtain the off-diagonal Hellmann-Feynman theorem \cite{Hellmann:1937,Feynman:1939}
\begin{equation}
    \langle \Psi_n | \partial_s \Psi_0 \rangle = - \frac{\langle \Psi_n | V | \Psi_0 \rangle} {\Delta_{n0}}.
    \label{eq:off-diag_H-F}
\end{equation}
Integrating Eq.~\eqref{eq:deriv_c} from $s=0$ to $s=1$ and using Eq.~\eqref{eq:off-diag_H-F}, we get
\begin{equation}
    c_n(1) = \int_0^1 ds \, \frac{\langle \Psi_n(s) | V | \Psi_0(s) \rangle}{\Delta_{n0}(s)} \exp\left( i T \int_0^s \Delta_{n0}(\tau) d\tau \right) + \mathcal{O}(T^{-2}).
    \label{eq:cn_integral}
\end{equation}

\subsubsection{Asymptotic Bound via Integration by Parts}
To extract the scaling behavior with respect to $T$, we perform integration by parts \cite{Kato:1950, Teufel:2003}. Let $\Theta_n(s) = \int_0^s \Delta_{n0}(\tau) d\tau$. We note that $\frac{d}{ds} e^{i T \Theta_n(s)} = i T \Delta_{n0}(s) e^{i T \Theta_n(s)}$, which implies
\begin{equation}
    e^{i T \Theta_n(s)} = \frac{1}{i T \Delta_{n0}(s)} \frac{d}{ds} \left( e^{i T \Theta_n(s)} \right).
\end{equation}
Substituting this into the integral in Eq.~\eqref{eq:cn_integral} gives
\begin{equation}
    c_n(1) = \int_0^1 ds \, \frac{\langle \Psi_n(s) | V | \Psi_0(s) \rangle}{\Delta_{n0}(s)} \left[ \frac{1}{i T \Delta_{n0}(s)} \frac{d}{ds} e^{i T \Theta_n(s)} \right].
\end{equation}
Integrating by parts yields
\begin{equation}
    c_n(1) = \left[ \frac{\langle \Psi_n(s) | V | \Psi_0(s) \rangle}{i T (\Delta_{n0}(s))^2} e^{i T \Theta_n(s)} \right]_0^1 - \frac{1}{i T} \int_0^1 ds \, e^{i T \Theta_n(s)} \frac{d}{ds} \left( \frac{\langle \Psi_n(s) | V | \Psi_0(s) \rangle}{(\Delta_{n0}(s))^2} \right).
\end{equation}
The first term represents the boundary contributions. The second term is an integral of a rapidly oscillating function with frequency proportional to $T$. Since the phase $\Theta_n(s)$ is strictly monotonic and the derivative term in the integrand is differentiable, the Riemann-Lebesgue lemma tells us that this remainder term scales as $\mathcal{O}(T^{-2})$ or smaller. Thus, the magnitude of the amplitude is dominated by the boundary terms
\begin{equation}
    |c_n(1)| \le \frac{1}{T} \left( \frac{|\langle \Psi_n(1)|V|\Psi_0(1)\rangle|}{(\Delta_{n0}(1))^2} + \frac{|\langle \Psi_n(0)|V|\Psi_0(0)\rangle|}{(\Delta_{n0}(0))^2} \right) + \mathcal{O}(T^{-2}).
\end{equation}

\subsubsection{Total Error and the Resolution Refinement Condition}
The total fidelity loss is the sum of the probabilities of finding the system in any excited state
\begin{equation}
    P_{\mathrm{loss}} = \sum_{n \neq 0} |c_n(1)|^2.
\end{equation}
Substituting the bound for $|c_n(1)|$ and applying the inequality $|a+b|^2 \le 2|a|^2 + 2|b|^2$ results in
\begin{equation}
    P_{\mathrm{loss}} \le \frac{2}{T^2} \sum_{n \neq 0} \left[ \frac{|\langle \Psi_n(1)|V|\Psi_0(1)\rangle|^2}{(\Delta_{n0}(1))^4} + \frac{|\langle \Psi_n(0)|V|\Psi_0(0)\rangle|^2}{(\Delta_{n0}(0))^4} \right] + \mathcal{O}(T^{-3}).
\end{equation}
From the definition of the minimum spectral gap, $\Delta E_{\rm min} \equiv \min_{s \in [0,1]} \Delta_{10}(s)$, we can use the inequality, 
\begin{equation}
 (\Delta_{n0}(s))^{-4} \le (\Delta E_{\rm min})^{-2} (\Delta_{n0}(s))^{-2},   
\end{equation} 
to factor out $(\Delta E_{\rm min})^{-2}$,
\begin{equation}
    P_{\mathrm{loss}} \le \frac{2}{T^2 (\Delta E_{\rm min})^2} \left( \sum_{n \neq 0} \frac{|\langle \Psi_n(1)|V|\Psi_0(1)\rangle|^2}{(\Delta_{n0}(1))^2} + \sum_{n \neq 0} \frac{|\langle \Psi_n(0)|V|\Psi_0(0)\rangle|^2}{(\Delta_{n0}(0))^2} \right) + \mathcal{O}(T^{-3}).
\end{equation}
The sums in parentheses correspond exactly to the definition of the instantaneous perturbative norm $\mathcal{N}_{\mathrm{pert}}(s)$ evaluated at $s=1$ and $s=0$. Invoking the condition that this norm is bounded above by $B$ gives
\begin{equation}
    P_{\mathrm{loss}} \le \frac{2}{T^2 (\Delta E_{\rm min})^2} (B + B) = \frac{4 B}{T^2 (\Delta E_{\rm min})^2} + \mathcal{O}(T^{-3}).
\end{equation}

\subsubsection{Resulting Estimate}
We require the final fidelity to be $\mathcal{F}(T) \ge 1 - \epsilon$, which corresponds to $P_{\mathrm{loss}} \le \epsilon$. Including the asymptotic error terms, we obtain the inequality
\begin{equation}
    \frac{4 B}{T^2 (\Delta E_{\rm min})^2} + \mathcal{O}(T^{-3}) \le \epsilon.
\end{equation}
Solving for the evolution time $T$, we obtain the sufficient time scaling
\begin{equation}
    T \sim \mathcal{O}\left( \frac{\sqrt{B}}{\sqrt{\epsilon} \Delta E_{\rm min}} \right).
\end{equation}
We observe that $T$ scales inversely with $\Delta E_{\rm min}$, inversely with $\sqrt{\epsilon}$, and linearly with $\sqrt{B}$.

We now consider how the bound $B$ scales with the number of particles $N$. If $V = H(1) - H(0)$ is an extensive operator with finite range, then the squared norm of the first-order wavefunction correction should scale linearly with system size, $B \sim \mathcal{O}(N)$.  We conclude that the adiabatic evolution time scales as
\begin{equation}
    T \sim \mathcal{O}\left( \frac{\sqrt{N}}{\sqrt{\epsilon} \Delta E_{\rm min}} \right). \label{eq:bound}
\end{equation}
This represents a significant advantage over general adiabatic bounds based on the global operator norm $\|\dot{H}\|$. Since $\|\dot{H}\| \sim \mathcal{O}(N)$, the general case requires an adiabatic time $T_{\rm general}$ with
\begin{equation}
    T_{\rm general } \sim \mathcal{O}\left( \frac{\|\dot{H}\|}{\sqrt{\epsilon} (\Delta E_{\rm min})^2} \right) \sim \mathcal{O}\left( \frac{N}{\sqrt{\epsilon} (\Delta E_{\rm min})^2} \right).
\end{equation}
\end{document}